\documentclass[a4paper,11pt]{article}
\pdfoutput=1
\usepackage{lineno}

\usepackage{jinstpub} 

\usepackage{caption}
\usepackage{subcaption}
\usepackage{graphicx} 
\usepackage{xcolor}
\usepackage{amsmath}
\usepackage[normalem]{ulem}
\DeclareMathOperator*{\argmin}{arg\,min}
\usepackage{mathtools}

\usepackage{braket}
\usepackage{subfiles}
\usepackage{placeins}

\title{A quantum algorithm for track reconstruction in the LHCb vertex detector}
\author[a]{D.~Nicotra,}
\author[a]{M.~Lucio~Martinez,}
\author[a]{J.~A.~de Vries,}
\author[a,b]{M.~Merk,}
\author[a]{K.~Driessens,}
\author[a]{R.~L.~Westra,}
\author[a]{D.~Dibenedetto}
\author[a]{and~D.~H.~C\'ampora~P\'erez}

\affiliation[a]{Universiteit Maastricht, Maastricht, The Netherlands}
\affiliation[b]{Nikhef National Institute for Subatomic Physics,  Amsterdam, The Netherlands}

\emailAdd{d.nicotra@maastrichtuniversity.nl}


\abstract{
High-energy physics is facing increasingly computational challenges in real-time event reconstruction for the near-future high-luminosity era. Using the LHCb vertex detector as a use-case, we explore a new algorithm for particle track reconstruction based on the minimisation of an Ising-like Hamiltonian with a linear algebra approach. The use of a classical matrix inversion technique results in tracking performance similar to the current state-of-the-art but with worse scaling complexity in time. To solve this problem, we also present an implementation as quantum algorithm, using the Harrow-Hassadim-Lloyd (HHL) algorithm: this approach can potentially provide an exponential speedup as a function of the number of input hits over its classical counterpart, in spite of limitations due to the well-known HHL Hamiltonian simulation and readout problems.
The findings presented in this paper shed light on the potential of leveraging quantum computing for real-time particle track reconstruction in high-energy physics.
}

\begin{document}

\maketitle
\flushbottom

\section{Introduction}

Progress in the field of experimental Particle Physics is often limited by statistical precision. For this reason, near-future upgrades to collider experiments, such as the Large Hadron Collider (LHC), are investing in an order of magnitude increase of data collection rates or \textit{instantaneous luminosity}. 
While this brings exciting physics targets within experimental reach~\cite{LHCb-U2,ATLASHLLHC,CMSHLLHC}, it comes with a drastic increase in the number of particles in the final state to be detected. These particles traverse the various sensor layers of the detector and generate \textit{hits}, which have to be reconstructed into the original particle trajectories in real-time for data storage selection and further processing.

As the complexity of particle track reconstruction scales with the number of sensor-hits to the power 2-3~\cite{searchbytriplet,DataAnaTechHEP}, depending on the algorithmic approach and geometry of the detector, time constraints and reducing fake rates provide challenges that are still to be overcome for High Luminosity-LHC (HL-LHC). Various approaches are under study, such as parallel reconstruction on GPUs~\cite{Allen} as well as the addition of picosecond timing information to sensors~\cite{VELOTimePix}. In this paper, we will explore the use of quantum algorithms for particle track reconstruction in the Upgrade LHCb vertex detector (VELO)~\cite{VELO}.

Quantum Computing (QC) represents one of the current frontiers of Computer Science and has seen rapid developments on both the hardware and the software sides over the past decade. QC has gained the attention of the High Energy Physics community~\cite{qc4hep}, offering the possibility of overcoming the current computational limitations of algorithms on simulation~\cite{QuantumSimulation1,QuantumSimulation2}, data analysis~\cite{QuantumDataAnalysis:btagging,QuantumDataAnalysis:qml1,QuantumDataAnalysis:qml2} and reconstruction~\cite{QuantumReconstruction}.

Several studies exist on the potential advantage of QC algorithms for track reconstruction. A  complexity study~\cite{ManganoEtAl}, shows some degree of potential speedup in quantum search algorithms using a seeding/following approach. Alternatively, a Quadratic Unconstrained Binary Optimisation (QUBO) formulation of the tracking problem and proof-of-principles have been built using quantum annealing~\cite{Bapst:2019llh} or a Variational Quantum Eigensolver (VQE)~\cite{QUBO2} to find solutions, obtaining reasonable tracking performance numbers but making no promise on HL-LHC conditions or timing improvements. Considering the reconstruction of tracks from hits as a clustering problem, one deals with a global algorithm, processing all the hits at once. An example of such an implementation was done at the LUXE experiment~\cite{LUXE}, where positrons trajectories have been reconstructed using a VQE to solve a QUBO problem, and a quantum Graph Neural Network (GNN)~\cite{crippa2023quantum,LUXEqTrack}. A similar approach, based on a hybrid quantum-classical GNN, has been tested on the TrackML dataset, which emulates HL-LHC conditions~\cite{qGraphs}. In both cases, a practical implementation on quantum hardware is limited by the number of input hits that the algorithm can process, due to the circuit becoming too deep to be compatible with current devices.

In this paper we adopt the QUBO formulation of the tracking problem, based on an custom Ising-like Hamiltonian, and we solve it in an alternative way, using a matrix inversion method. The well-known Harrow-Hassadim-Lloyd (HHL) algorithm~\cite{HHL} is employed for matrix inversion of a Hamiltonian embedded in a quantum circuit. We show that the preparation of the input quantum state can be executed efficiently on a quantum device and that the matrix to be inverted is highly sparse and well-conditioned, matching the requirements of HHL. We benchmark our algorithm on LHCb simulated data and test a quantum implementation on toy model events. When the hypotheses for an efficient implementation of HHL will be met, this approach promises an exponential speedup over classical counterparts.

\section{Track reconstruction in LHCb vertex detector}

\subsection{The LHCb vertex detector}
\begin{figure}[ht]
    \centering
    \includegraphics{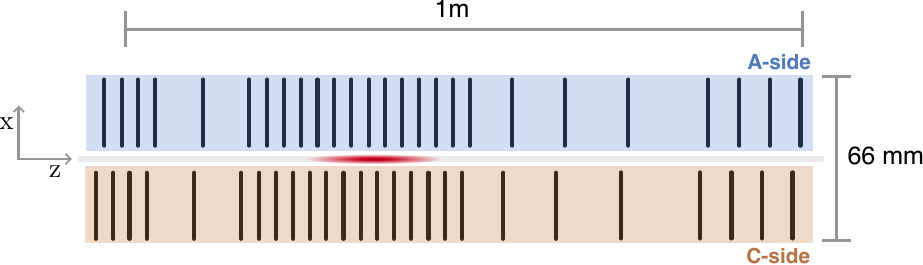}\\
    \includegraphics{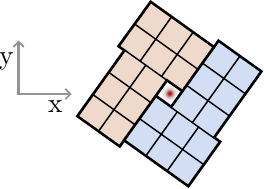}
    \caption{Schematic representation of the VELO. The 52 detection modules are arranged on the so-called A-side (left side) and C-side (right side). The interaction region is located in the area with the highest density of modules and depicted in red.}
    \label{fig:VELO}
\end{figure}

The LHCb detector~\cite{LHCb1,LHCb2} is a single-arm forward spectrometer that covers the pseudo-rapidity\footnote{The pseudo-rapidity is defined as $\eta = - \log{\tan \frac{\theta}{2}}$, where $\theta$ is the angle between the three-momentum of the particle and the beam axis.} range $2 < \eta < 5$. The reconstruction use case investigated is based on the LHCb VELO tracker~\cite{VELO}. The VELO consists of a series of vertically oriented pixel sensors arranged along the beam-line (on the $z$-axis), positioned close to the collision point of the LHC, and measuring the $xy$ coordinates of passing particles (see Figure \ref{fig:VELO}). The collisions occur in a 5 cm long region along the $z$-axis, with detection layers surrounding the beam-line axis $x=y=0$.
The VELO is placed upstream to the dipole magnet, in a region practically free from magnetic field, leading to straight tracks and simplifying the track model when considering only hits in the VELO.
\subsection{Track reconstruction algorithms}
The reconstruction of particle trajectories in high-energy physics experiments at a hadron collider is a computing-intensive task due to the high multiplicity of particles in the final state.

Approaches to track reconstruction can be  classified into two categories: \textit{local} and \textit{global} algorithms. Global algorithms process all the hits in an event simultaneously, aiming to provide a globally optimal solution to the track-finding problem; this category includes approaches based on the Hough transform methods~\cite{TrackingHough1}, Hopfield neural networks~\cite{TrackingHopfield1,TrackingHopfield2}, etc.
On the other hand, local algorithms process the hits in a sequential way, starting from a set of \textit{track seeds}: as a consequence, the result of the algorithm may depend on the choice of the starting seeds and processing order of the hits. This class includes Kalman filter techniques~\cite{TrackingKalman1,TrackingKalman2}.

The choice of which algorithm fits best with a specific use-case depends on the specifics of the detector, the operating conditions and the available computing resources. Generally, global methods tend to scale better in time complexity, while for space complexity local algorithms may be more suitable. Both types of methods have been widely adopted in particle physics experiments~\cite{TrackingHopfield2,TrackRecoReview2}.

The pattern recognition problem used to find particle trajectories is increasingly approached using a heterogeneous computing model. Where before the task was fully CPU-based, recent developments add GPGPU-based algorithms in LHCb.

The current state-of-the-art VELO track reconstruction algorithm is such a GPGPU-based algorithm called Search-by-Triplet~\cite{searchbytriplet} exploiting a high degree of parallelism. Track candidates are built from hits in three consecutive layers close to the primary proton-proton interaction point, which are then projected  in $z$ to find hits in the following detector layers (\textit{track following}). The quantum algorithm investigated in this paper, instead, is a global method considering all hits at once, and the performance of Search-by-Triplet will act as a baseline comparison.

\section{Tracking with Ising-like Hamiltonian}

\subsection{Ising-based methods}
In this paper, we present a quantum algorithm for track reconstruction based on the minimisation of an Ising-like Hamiltonian. The algorithm takes, as an input, a collection of hits left by a set of particles over the several layers of the detector. The algorithm will reconstruct each particle trajectory as a collection of track segments (also called doublets) that belong to that track. The first step consists in constructing all the possible doublets of hits in consecutive modules of the detector: each doublet is associated with a binary variable $S_i \in \{0,1\}$ where

 \begin{equation*}
     S_i = \begin{cases} 1  & \text{if the doublet is part of a track}\\ 0 & \text{otherwise}\end{cases} \,.
 \end{equation*}
 
 \noindent Therefore, the track reconstruction task consists of determining the vector $\mathbf S=\{S_1, S_2, ..., S_N\}$ containing the correct activation state of every doublet, i.e. $1$ if it is part of any track or $0$ otherwise.
\begin{figure}[ht]
    \centering
    \begin{subfigure}{\textwidth}
    \centering
        \includegraphics[width=.75\textwidth]{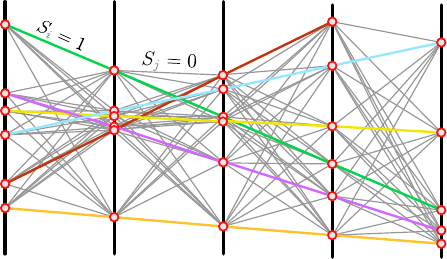}
        \caption{}
        \vspace{.5cm}
        \label{fig:all_doublets}
    \end{subfigure}
    \begin{subfigure}{.48\textwidth}
    \centering
        \includegraphics[width=\textwidth]{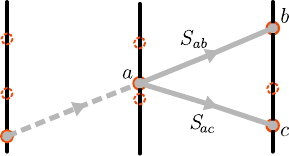}
        \caption{}
        \label{fig:bifurcations}
    \end{subfigure}
    \hspace{.2cm}
    \begin{subfigure}{.48\textwidth}
    \centering
        \includegraphics[width=\textwidth]{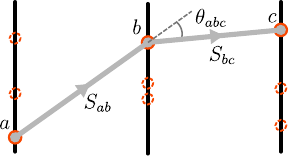}
        \caption{}
        \label{fig:angle}
    \end{subfigure}
    
    \caption{(a) Event display with all the possible doublets represented: the ones that are part of a real trajectory are highlighted while the others are left in grey. (b) Diagram showing an example of a track bifurcation. (c) Diagram explicitly representing the angle $\theta_{abc}$ between two adjacent doublets $S_{ab}$ and $S_{bc}$.}
    \label{fig:diagrams}
\end{figure}
Ising-based approaches to this problem consist in finding a suitable quadratic Hamiltonian in the form

 \begin{equation}
 \label{eqn:ising-like-hamiltonian}
     \mathcal{H}(\mathbf{S}) = -\frac{1}{2}\sum_{i,j}A_{ij}S_i S_j + \sum_i b_i S_i = -\frac{1}{2} \mathbf{S}^\mathrm{T} A \mathbf{S} + \mathbf{b}^\mathrm{T} \mathbf{S}\,,
 \end{equation}
 such that the solution to the problem coincides with its ground state. This represents a generalised Ising Hamiltonian with couplings described by the real symmetric matrix $A$ (the Ising interaction matrix) and an external field described by the real vector $\mathbf b$ (bias vector). In general, finding the optimal value of $\mathbf{S}$ is a QUBO problem and requires a number of computations of $\mathcal{H}$ that scales exponentially with the size of $\mathbf{S}$~\cite{Bapst:2019llh}. The explicit representation of \eqref{eqn:ising-like-hamiltonian}, i.e. the coupling matrix $A$ and bias vector $\mathbf{b}$, must be chosen in such a way that the minimum of $\mathcal{H}$ corresponds to the solution of the tracking problem: in this paper, we adopt a modified version of the Denby-Peterson Hamiltonian.

\subsection{Denby-Peterson Hamiltonian} The Denby-Peterson (DP) Hamiltonian~\cite{Denby,Peterson} is an Ising-like Hamiltonian in the form \eqref{eqn:ising-like-hamiltonian}, which is often used in track reconstruction algorithms. It favours the reconstruction of smooth and long tracks while preventing non-physical bifurcations. Let $S_{ab}$ be the oriented segment stemming from hit $a$ to hit $b$. The DP Hamiltonian can be expressed in the form

\begin{align}
\label{eqn:dp-hamiltonian}
 \mathcal{H}_\mathrm{DP}= \mathcal{H}_\mathrm{ang}(\mathbf{S, \lambda})
 + \alpha\mathcal{H}_\mathrm{bif}(\mathbf{S})
 + \beta\mathcal{H}_\mathrm{occ}(\mathbf{S},N_\mathrm{hits})\,,
 \end{align}
highlighting the following three main components:
\begin{description}
    \item[Angular term]
    
    \begin{align}
    \label{eqn:dp-angular-term}
        \mathcal{H}_\mathrm{ang} = -\frac{1}{2}\sum_{a,b,c} \left(\frac{\cos^\lambda \theta_{abc}}{r_{ab} + r_{bc}} S_{ab}S_{bc}\right)\,,
    \end{align}
    where $\theta_{abc}$ is the angle between the segments $S_{ab}$ and $S_{bc}$, as depicted in Figure \ref{fig:angle}; $r_{ab}$ and $r_{bc}$ are the length of the segments $S_{ab}$ and $S_{bc}$ respectively. This term favours aligned triplets (i.e. where $\theta_{abc}$ is small) with short doublets. $\lambda$ is a constant integer: large values of $\lambda$  penalise less straight triplets in favour of aligned ones. 

    \item [Bifurcation term]
    
    \begin{equation}
   \label{eqn:dp-bifurcation-term}
         \mathcal{H}_\mathrm{bif} = +\frac{1}{2} \left( \sum_{b \ne c}S_{ab}S_{ac} + \sum_{a \ne c} S_{ab}S_{cb}\right)
    \end{equation}
   
    is weighted by the Lagrangian multiplier $\alpha \ge 0$. This term penalises, with a constant penalty $+\alpha$, bifurcations in the trajectories like the ones shown in Figure \ref{fig:bifurcations}.

    \item[Occupancy term]
    
    \begin{equation}
        \label{eqn:dp-occupancy-term}
        \mathcal{H}_\mathrm{occ} = +\frac{1}{2}\left( \sum_{a,b}S_{ab} - N_\mathrm{hits} \right)^2\,,
    \end{equation}
    where $N_\mathrm{hits}$ is the number of hits in the detector. This term constrains the number of active doublets to be roughly equal to the number of hits in the event and it is weighted by the Lagrangian multiplier $\beta \ge 0$.
\end{description}
The constants $\lambda$, $\alpha$ and $\beta$ are free parameters of the model and must be fixed in advance according to an appropriate hyper-parameter search.

\begin{figure}
    \centering
    \includegraphics[width=0.8\textwidth]{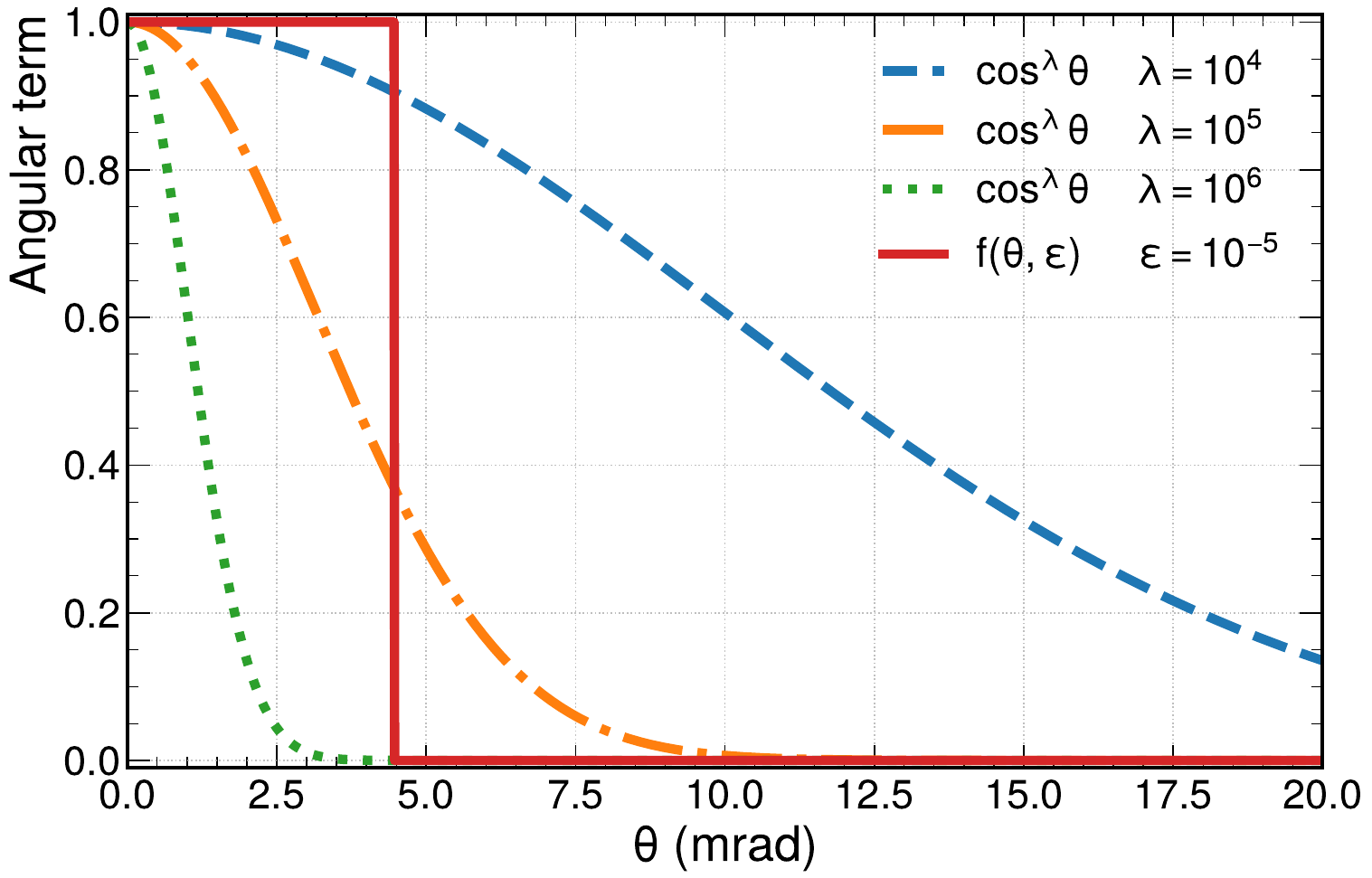}
    \caption{Plot of $\cos^\lambda(\theta)$ for several values of $\lambda$, compared with the step function $f(\theta, \varepsilon)$}
    \label{fig:cosine_and_f}
\end{figure}

The DP Hamiltonian has been designed to reconstruct curved trajectories, minimising the amount of bifurcations in the reconstructed tracks. However, as the LHCb VELO operates in a region practically free of magnetic field, tracks are approximately straight lines, therefore we expect the angular term to have the dominant contribution, relatively to the other two terms, which can be eventually neglected. Moreover, as the quantum implementation for this algorithm is expected to scale better when number of non-zero couplings in the Hamiltonian is small, it can benefit from a simplification of the terms described above. For these reasons, in this work, we use a modified version of the DP Hamiltonian, removing the bifurcation and occupancy terms (i.e. $\alpha$ and $\beta$ are fixed to 0), and using a simplified angular term:

\begin{equation}
    \mathcal{H}(\mathbf{S}) = \mathcal{H}_{\rm ang}(\mathbf{S}, \varepsilon)\,,
    \label{eqn:ourHamiltonian}
\end{equation}
with

\begin{align}
    \mathcal{H}_\mathrm{ang}(\mathbf{S, \varepsilon}) = -\frac{1}{2}\sum_{a,b,c}f(\theta_{abc}, \varepsilon)S_{ab}S_{bc} && 
        f(\theta_{abc}, \varepsilon) = \begin{cases}
            1 & \text{if } \cos \theta_{abc} \ge 1 -\varepsilon \\
            0 & \text{otherwise}
        \end{cases}\,.
        \label{eqn:mod_angular}
    \end{align}
This term, compared to \eqref{eqn:dp-angular-term}, removes the dependence on the segment length and replaces the smooth cosine behaviour with a step function of width $\arccos(1 -\varepsilon)$. The constant $\varepsilon$ represents an angular tolerance for the model: only consecutive segments that form an angle smaller than $\arccos(1 -\varepsilon)$ are coupled by the angular term, while the others are left uncoupled. The parameter $\varepsilon$ allows to deal with trajectories that are not perfectly straight. Therefore, its numerical value can be related to the physical characteristics of the detector, such as its geometry and resolution. It also allows taking into account multiple-scattering effects, i.e. when a particle is slightly scattered by the interactions with the detector material. Figure \ref{fig:cosine_and_f} shows the decay behaviour of $\cos^\lambda(\theta)$ for different values of the exponent $\lambda$, compared to the step function $f$ with $\varepsilon = 10^{-5}$.
   

\section{The quantum algorithm}

The DP Hamiltonian has been employed in track reconstruction algorithms based on the optimisation of Hopfield networks~\cite{Denby,Peterson,TrackingHopfield1,TrackingHopfield2}: nevertheless, these approaches become unfeasible in high hit multiplicity conditions due to the scaling of the size of the network. Several quantum versions of the Hopfield network are available in the literature, trying to exploit the unique features of quantum computation to achieve an advantage~\cite{QuantumHopfield1,QuantumHopfield2,QuantumHopfield3,QuantumHopfield4,Rebentrost}.
Our quantum algorithm follows a similar approach to Rebentrost et al.~\cite{Rebentrost}, in which the optimisation of a Hopfield network is turned into a matrix inversion problem and solved with the Harrow-Hassadim-Lloyd (HHL) algorithm (also known as quantum algorithm for linear systems of equations), which has the potential of achieving an exponential advantage with respect to its classical counterpart~\cite{HHL}. 

\subsection{Hamiltonian Relaxation and local minimum} 
The first step consists of relaxing the original problem, where we allow the binary-valued spin $S_i$ in the Ising-like Hamiltonian \eqref{eqn:ising-like-hamiltonian} to have any possible real value $S_i \in \mathbb{R}$, instead of only 0 or 1. This way, the Hamiltonian is a differentiable function of the variables $S_i$, and this allows us to calculate the gradient. The drawback of this approach is that we allow each $S_i$ to have a non-physical state, such as a negative or fractional value: to solve this problem, a threshold operation is performed in the final stage of the algorithm, which will ensure valid final states. Using the fact that $A$ is symmetric, the gradient of $\mathcal{H}$ with respect to $\mathbf{S}$ has the following expression

\begin{equation}
\label{eqn:hamiltonian-gradient}
    \nabla \mathcal{H}= - A \mathbf{S} + \mathbf{b} \,.
\end{equation}
Now we look for a $\mathbf{S}$ that satisfies $\nabla \mathcal{H} (\mathbf{S}) = 0$, the necessary condition for being a local minimum. This implies solving the square system of linear equations

\begin{equation}
\label{eqn:system-of-linear-equation}
    A\mathbf{S} = \mathbf{b}\,.
\end{equation}
In general a solution for \eqref{eqn:system-of-linear-equation} is not guaranteed to exist (i.e. if $\det A = 0$). To overcome this, we look for a least-squares solution for this system:

\begin{equation}
\label{eqn:least-square}
    \Tilde{\mathbf{S}} = \argmin_\mathbf{S}||A\mathbf{S} - \mathbf{b}||_2\,,
\end{equation}
where $||\cdot||_2$ denotes the Euclidean 2-norm. In general, \eqref{eqn:least-square} may have multiple solutions: in that case, we select the one with the smallest 2-norm. This problem has a unique solution that is provided by the Moore-Penrose pseudo-inverse matrix~\cite{MoorePenrose} $A^+$ applied to the vector $\mathbf{b}$:

\begin{equation}
    \Tilde{\mathbf{S}} = A^+\mathbf{b}\,.
\end{equation}
If $\det A \ne 0$ then the pseudo-inverse $A^+$ and the inverse $A^{-1}$ coincide.
\subsection{Regularization terms}
In order to deal with the relaxed version of the problem, we have to introduce two regularisation terms in the Hamiltonian \eqref{eqn:ourHamiltonian}, which becomes

\begin{equation}
    \mathcal{H}(\mathbf{S}) = \mathcal{H}_\mathrm{ang}(\mathbf{S, \varepsilon}) + \gamma\mathcal{H}_\mathrm{spec}(\mathbf{S}) + \delta\mathcal{H}_\mathrm{gap}(\mathbf{S})\,:
\end{equation}
\begin{description}
    \item [Spectral term]
    
    \begin{align}
        \mathcal{H}_\mathrm{spec} = \frac{1}{2} \sum_{ab}S_{ab}^2 = \frac{1}{2}\mathbf{S}^\mathrm{T} \mathbb{I} \mathbf{S}
        \label{eqn:spectralterm}
    \end{align}
    This term is purely quadratic and contributes to the diagonal of the matrix $A$. The purpose of this term is to shift the eigenvalues spectrum of $A$ by a constant $\gamma$ to make it positive-definite and avoid numerical instabilities. However, $\gamma$ cannot be chosen arbitrarily large as this would result in a polynomial increase of the run-time~\cite{Rebentrost}.
    \item[Gap term]
    
    \begin{equation}
        \mathcal{H}_\mathrm{gap} = \frac{1}{2}\sum_{ab}(1-2S_{ab})^2 = \frac{1}{2}\left[\sum_{ab}1 + 4\sum_{ab}S_{ab}^2 - 4\sum_{ab}S_{ab}
        \right]
        \label{eqn:gapterm}
    \end{equation}
    Due to the presence of linear and quadratic contributions, this term contributes to both the matrix $A$ and the vector $\mathbf{b}$, by a constant $\delta$. The purpose of this term is to enforce a gap in the spectrum of the solution vector $\Tilde{\mathbf{S}}$, suppressing the occurrence of ambiguous states that are not close to 0 or 1. This allows to identify a stable threshold value $T$ and discretise the output values as
    
    \begin{equation}
        \Tilde{S}_i \to 
        \begin{cases}
            \text{on} & \text{if } \Tilde{S}_i > T\\
            \text{off} & \text{if } \Tilde{S}_i \le T\\
        \end{cases}
    \end{equation}
\end{description}

The numerical values of the hyper-parameters have been chosen according to the testing conditions of the algorithm. The parameter $\varepsilon$, which appears in the angular term \eqref{eqn:mod_angular}, has been fixed to match the average multiple scattering angle in a LHCb VELO module~\cite{VELOScattering}. The spectral parameter $\gamma$ has been fixed such that the eigenvalues spectrum of $A$ stays positive. It has been observed that the value of gap parameter $\delta$ is correlated with the threshold $T$, but does not directly affect the performance of the algorithm: therefore $\delta$ has been fixed to $1.0$ and the corresponding threshold $T$ has been determined as the average gap midpoint observed on events generated with a toy model, described in Section~\ref{sect:toymodel}.
The numerical values of all hyper-parameters are reported in the table below.
\begin{center}
    \begin{tabular}{|c|c|c|c|}
\hline
\textbf{Angular} & \textbf{Spectral} & \textbf{Gap} & \textbf{Threshold} \\
$\varepsilon$         & $\gamma$               & $\delta$          & $T$ \\ \hline
$10^{-5}$             & $2.0$                  & $1.0$             & $0.45$\\ \hline
\end{tabular}
\end{center}

\subsection{Solving the system using quantum computing}
\begin{figure}[ht]
    \centering
    \includegraphics[width=0.7\textwidth]{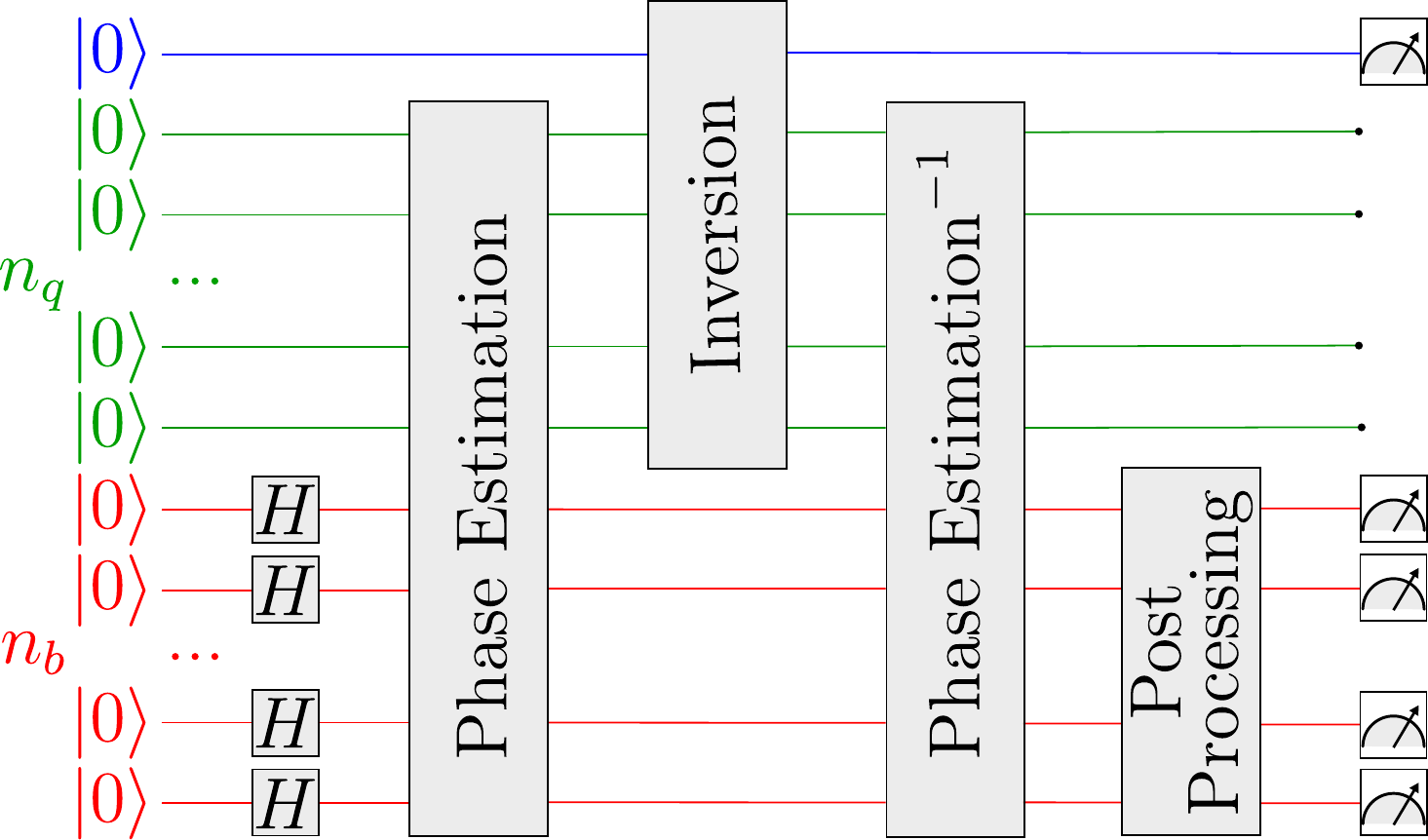}
    \caption{Quantum circuit that implements the algorithm. In red, the $\ket{\Phi}_b$, which encodes the $\mathbf{b}$ input vector and the $\mathbf{S}$ output vector. In green, the $n_q$-qubit ancilla register used for QPE. In blue, the additional ancilla qubit used for the matrix inversion. }
    \label{fig:circuit}
\end{figure}

The next step consists of solving the linear system \eqref{eqn:system-of-linear-equation} where the coupling matrix $A$ obtains contributions from each term in the Hamiltonian, including the hyper-parameters

$$A = A_\mathrm{ang} + A_\mathrm{spec} + A_\mathrm{gap}\,,$$
while the only non-zero contribution to the bias vector $\mathbf b$ comes from the gap term

$$\mathbf{b} = \mathbf{b}_\mathrm{gap} = \delta(1,1,1, ...,1)\,.$$
To solve the system of linear equations~\eqref{eqn:system-of-linear-equation} we use the HHL quantum algorithm~\cite{HHL,Rebentrost}. Under proper assumptions, the HHL algorithm is able to solve a $N \times N$ sparse linear system of equations in $\mathcal{O}(\kappa^2\log N)$ run-time, where $\kappa$ is the condition number of the matrix~\cite{numlinalg}, achieving an exponential advantage in the problem size over classical algorithms~\cite{HHL}. This advantage is achieved thanks to a powerful subroutine called Quantum Phase Estimation (QPE), which is also used in other quantum algorithms that show an exponential advantage, such as Shor's algorithm~\cite{Shor}.  It requires the matrix $A$ to be square and its dimension to be a power of 2. If that is not the case, in practice, it is always possible to pad $A$ to round up its dimension to the closest power of 2. Similarly, the $\mathbf{b}$ vector can always be padded with a suitable number of ones and normalised.
\begin{linenomath}
\begin{align*}
    A \to \begin{pmatrix}
    A & 0 \\
    0 & \mathbb{I}
\end{pmatrix} &&
\mathbf{b} \to \frac{1}{C}(\mathbf{b}, 1, 1 ..., 1) \,.
\end{align*}
\end{linenomath}
Therefore we will henceforth assume $N$ to be a power of 2 and $||\mathbf{b}|| = 1$. QPE also requires the matrix $A$ to be sparse, in order to be implemented efficiently. The Hamiltonian presented in this work has been designed to match the sparsity condition in practice: a more detailed study about the sparsity dependence on the event size can be found in Appendix~\ref{sect:sparsity}. The HHL run-time complexity has a quadratic dependence on the condition number $\kappa$: a study of the behaviour of the condition number with respect to the number of particles and detection layers has revealed that $\kappa$ can be upper-bounded irrespective of the problem size and therefore does not degrade the exponential advantage of HHL. More details can be found in Appendix~\ref{sect:kappa}. The algorithm requires a register of $n_b = \log_2 N$ qubits $\ket{\Phi}_b$ for storing the vector $\mathbf{b}$ and the solution vector $\mathbf{S}$. An additional register $\ket{\Phi}_q$ of $n_q$ qubits is required for QPE, according to the required precision on the estimated eigenvalues: $n_q$ is chosen as follows

$$
n_q = \max(1 + n_b, \lceil\log_2(\kappa + 1)\rceil)\, .
$$
However, since $\kappa$ is upper-bounded, $1 + n_b$ is dominant for $N \geq 4$.
Finally, the matrix inversion step requires an additional single-qubit register $\ket{\Phi}_i$. The global quantum state is represented as the tensor product of these 3 registers

$$
\ket{\Psi}= \ket{\Phi}_i\ket{\Phi}_q\ket{\Phi}_b\,.
$$
The algorithm is implemented by the quantum circuit shown in Figure \ref{fig:circuit} and consists of the following steps:
\begin{enumerate}
    \item \textbf{Initialisation}: Each qubit is initialised to the $\ket{0}$ computational state
    
    $$
        \ket{\Psi}= \ket{0}_i\ket{0}_q\ket{0}_b\,.
    $$
    \item \textbf{State preparation}: the entries $b_i$ of the vector $\mathbf{b}$ have to be loaded into the amplitudes of a quantum register $\ket{\Phi}_b$.
    
    $$
    \ket{\Phi}_b = \ket{b} = \sum_{i = 1}^N b_i \ket{i}\,,
    $$
    where $\ket{i}$ represents the $i$-th computational basis vector. This requires a suitable unitary $U_b$ to construct the state $\ket{b}$ from the initial state $\ket{0}$, namely $\ket{b} = U_b \ket{0}$. This is usually a strong limitation of the HHL algorithm as no efficient arbitrary state preparation is currently known. In this application, the vector $\mathbf{b}$ gets its only contribution from the gap term\eqref{eqn:gapterm}, which is a constant. Therefore, including the normalisation constant $C$, $\mathbf{b}$ has the following structure
    
    $$
     \mathbf{b} = \frac{1}{C}(1, 1, 1, ... ,1)\,.
    $$
    The corresponding state $\ket{b}$ can be constructed from the initial state $\ket{0}$ by applying a single Hadamard gate to each of the $n_b$ qubits.
    \item \textbf{Quantum Phase Estimation}:
    The QPE subroutine is applied to the operator $U = e^{iA}$ and will determine its phases with $n_q$-bit precision, where $n_q$ is the number of ancilla qubits. QPE employs multiple controlled executions of integer powers of $U$, $U^2, U^3, ...$. In order for this to be done efficiently, one needs a quantum circuit that implements the Hamiltonian simulation operator $U(t) = e^{iAt}$. Such an efficient implementation for this approach is not available yet and therefore represents the current limitation of this algorithm.

    \item \textbf{Matrix inversion}: The matrix $A$ is inverted in its own eigenbasis, by taking the reciprocal of each eigenvalue. This is accomplished by a controlled rotation of the ancilla qubit $\ket{\Phi}_i$. The algorithm ensures the successful inversion of the matrix if the ancilla is measured in the state $\ket{1}$. If $\ket{\Phi}_i$ is found to be in state $\ket{0}$, then the state is discarded and steps 1,2,3 and 4 are repeated until $\ket{1}$ is measured.
    \item \textbf{Uncomputing}: The ancilla registers $\ket{\Phi}_q$ are cleaned up by uncomputing them: in practice, the QPE subroutine is inverted and applied to the $\ket{\Phi}_q$ register to bring them back to their initial state. This step is necessary to avoid that, after the computation, the floating state of the ancilla registers interferes with the measurement of the qubits where the result is stored~\cite{Uncomputinmg}.

    \item \textbf{Post-processing and measurement}: At this stage, the register $\ket{\Phi}_b$ contains the entries of the solution vector
    
    $$
        \ket{\Phi}_b = \ket{S} = \sum_{i=1}^N S_i \ket{i}\,.
    $$
    The measurement of all the amplitudes of the final state, at this point, would require at least $N$ repetitions of the algorithm, voiding the exponential advantage. One can employ quantum post-processing techniques on the final state and measure the expectation value of an observable $\braket{S|\mathcal{O}|S}$. The identification of such an observable is currently out of the scope of this work, and state-vector analyses have been performed using a quantum simulator. The quantum algorithm has been implemented~\cite{codeZenodo} using the Qiskit open-source Python library~\cite{Qiskit}.
\end{enumerate}

\section{Results}

The algorithm is validated by assessing the track reconstruction performance using simulated events. Due to the lack of an efficient Hamiltonian simulation subroutine, processing LHCb simulated events using HHL is currently unfeasible: the HHL implementation available inside the Qiskit library produces circuits that are too deep to be run on current quantum hardware and to be simulated efficiently. For this reason, we have opted for a hybrid validation procedure: track reconstruction performances have been measured on LHCb simulated events, replacing the HHL algorithm with a classical solver, while the quantum approach has been tested using much smaller and cleaner events generated with a toy model. In the full simulations, the LHCb VELO geometry is simulated including multiple scattering of the particles on the detector material, as well as realistic hit detection efficiency and resolutions.
\subsection{Detector toy model}
\label{sect:toymodel}
Toy simulations are used for R\&D purposes of the algorithm: they implement a flexible detector geometry and allow us to study the algorithm in ideal conditions, i.e. perfectly straight tracks and no spurious hits, varying the number of detection layers and particle multiplicity. The detector is made of a tunable number of layers placed parallel to the $xy$-plane at regular distances from the origin. Particle trajectories are lines stemming from the origin, generated uniformly within the geometrical acceptance of the detector modules. Hits are collected from the intersections between the track and the detector layers. For the toy simulations ideal detector performance, in particular, no scattering, 100\% detection efficiency and perfect hit resolutions are assumed. The code implementing the detector toy model is available in Ref.~\cite{codeZenodo}.
\subsection{Track reconstruction performance on LHCb simulated data}
\begin{figure}[ht]
    \centering
    \includegraphics[width=\textwidth]{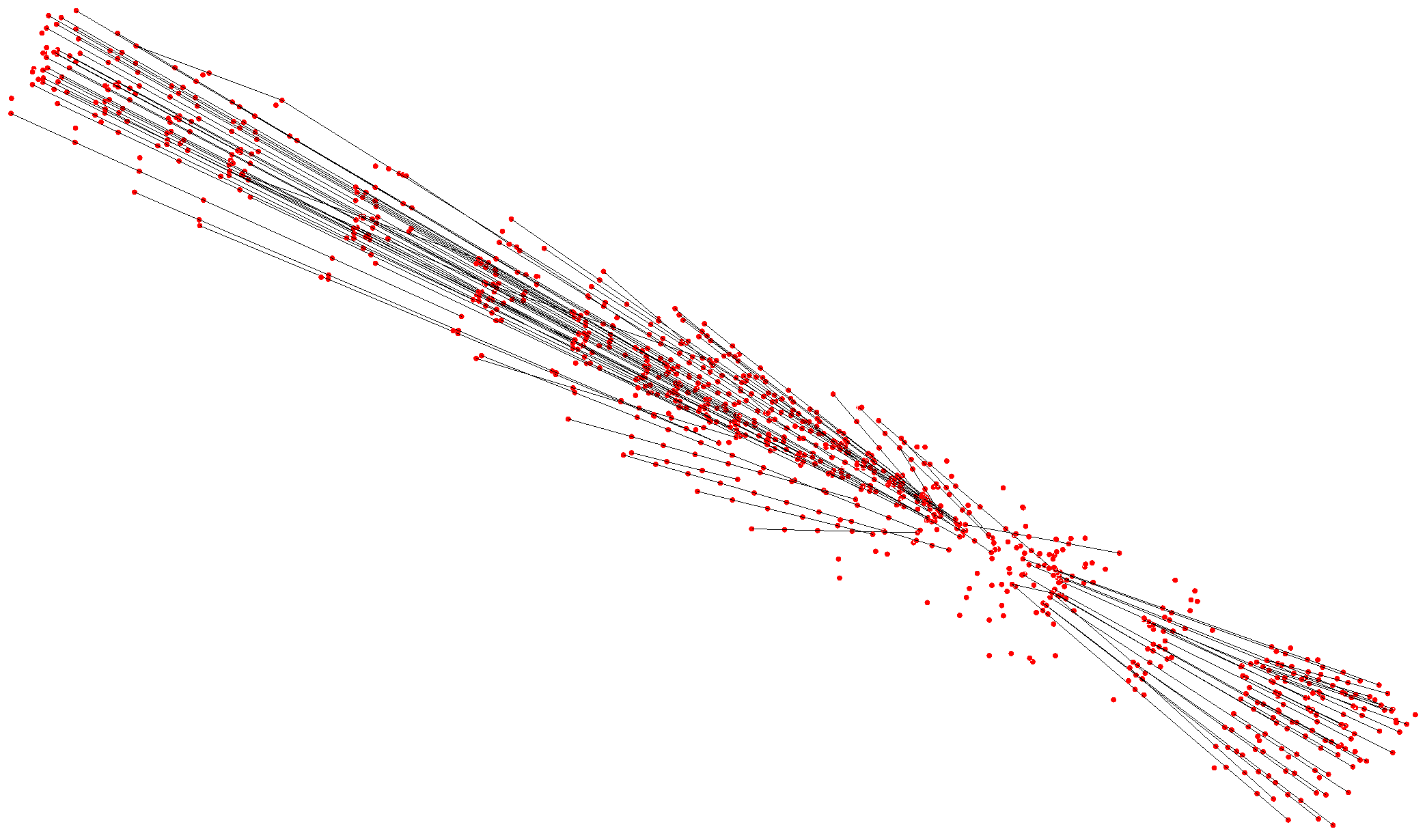}
    \caption{Event display of the hits (in red) coming from a collision event in half of the VELO together with the tracks reconstructed using the classical linear Ising solver presented in this work (in black).}
    \label{fig:velo_evdisp}
\end{figure}

\begin{figure}[ht]
    \centering
    \includegraphics[height=0.3\textheight]{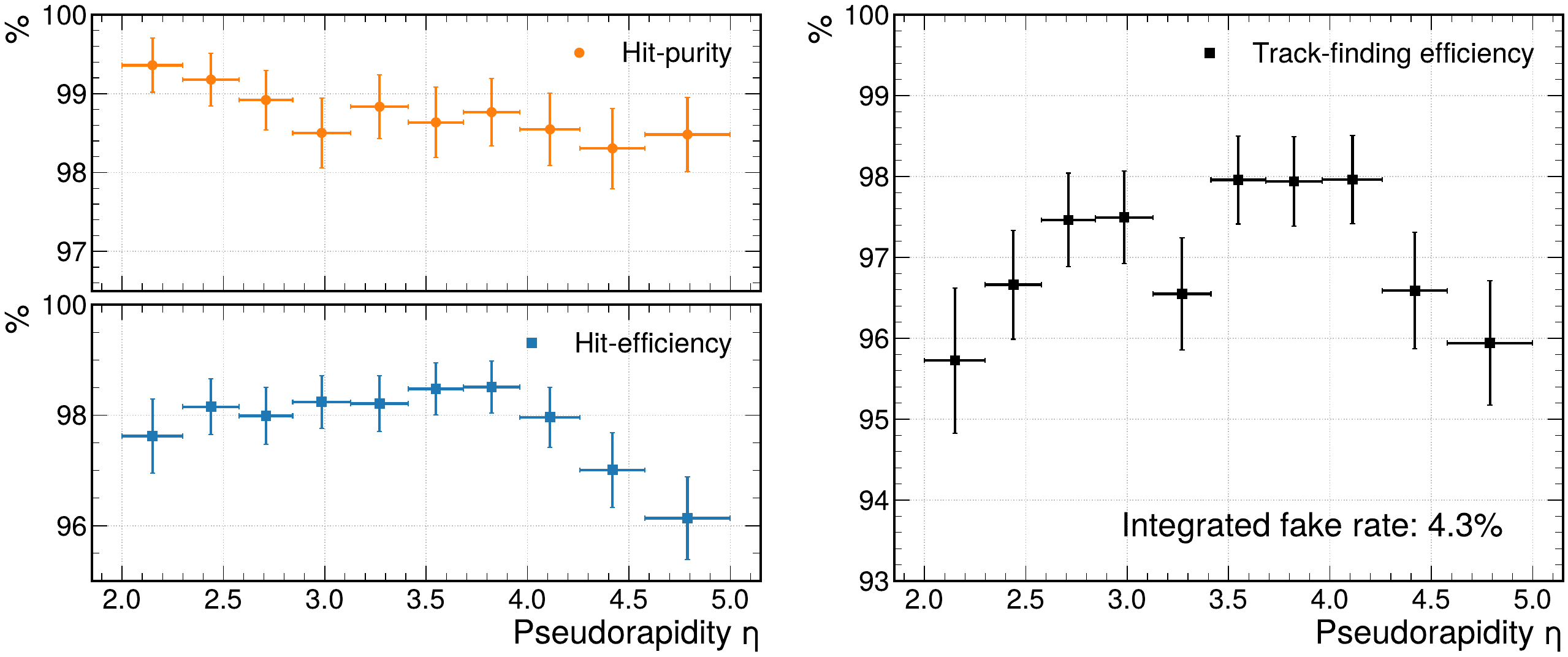}\\
    \includegraphics[height=0.3\textheight]{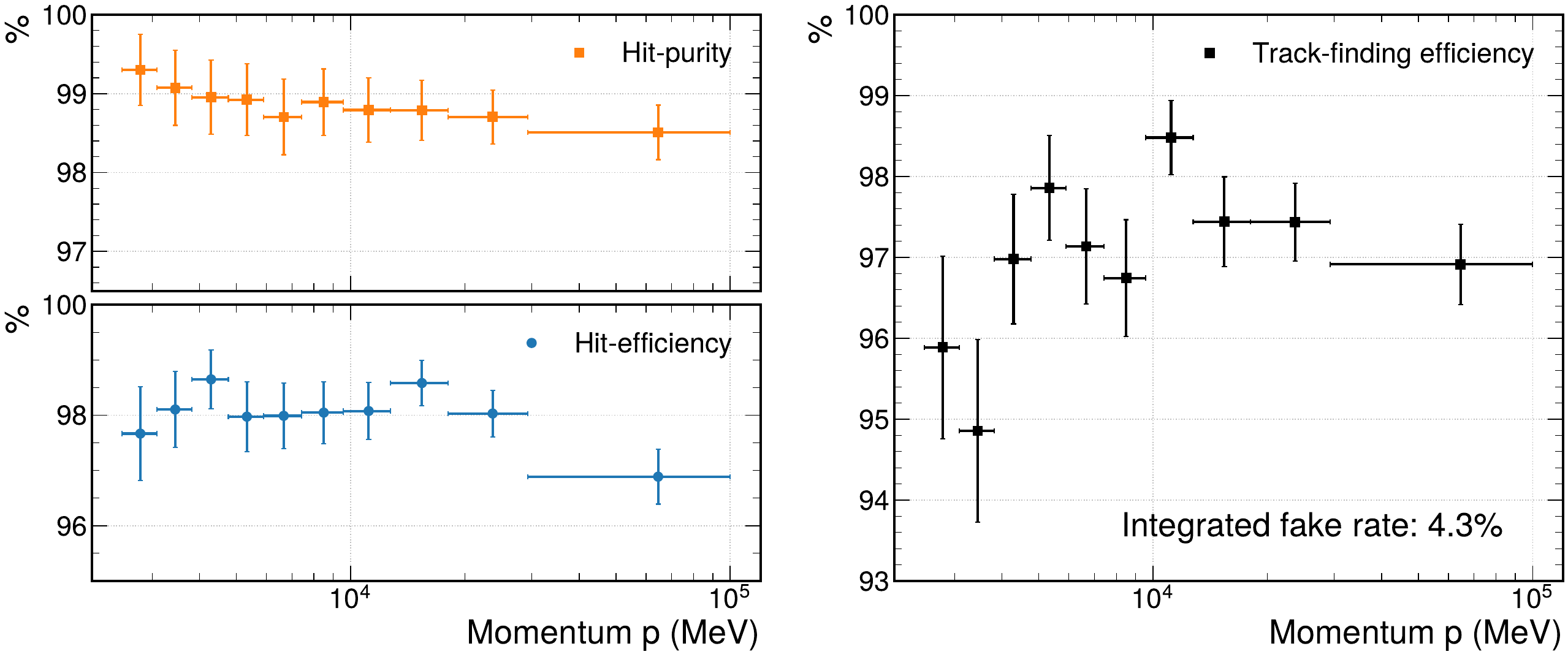}\\
    \includegraphics[height=0.3\textheight]{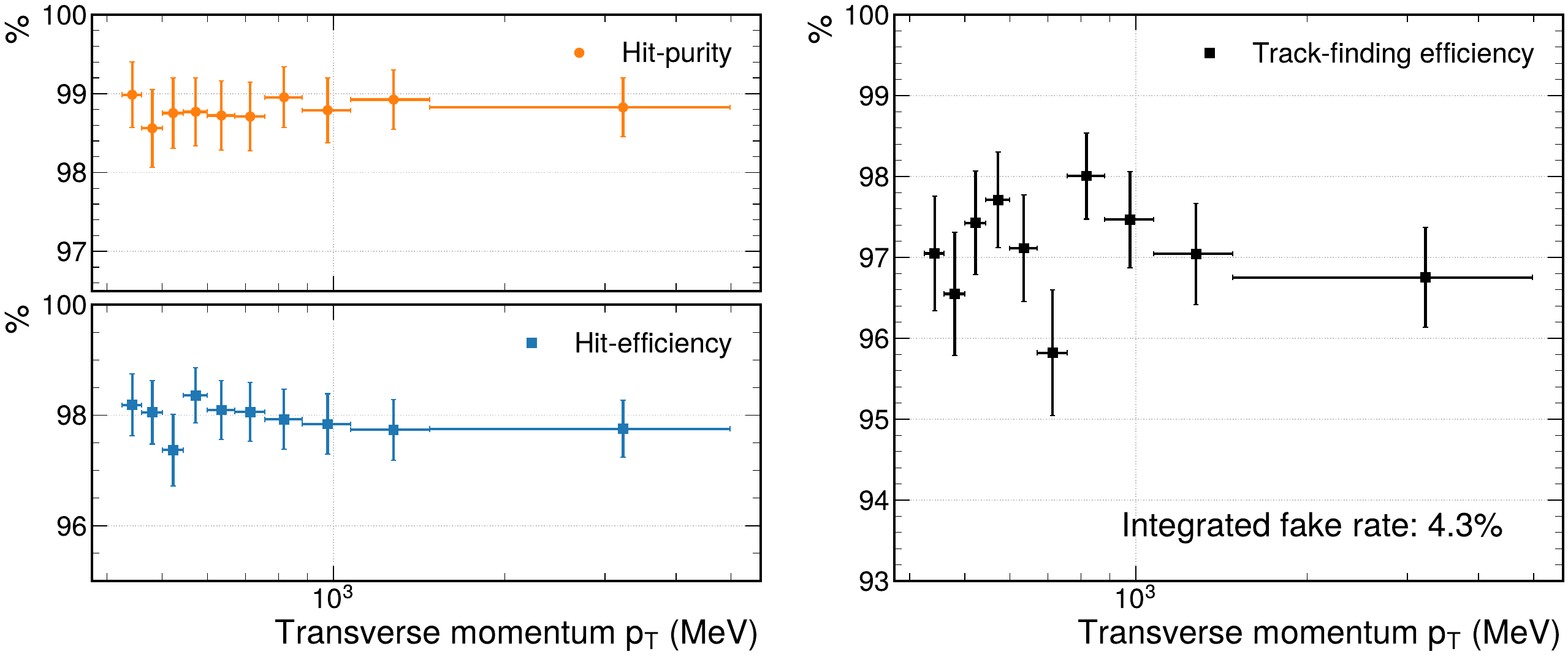}
    \caption{Track reconstruction performances in bins of $\eta$, $p$ and $p_\mathrm{T}$}
    \label{fig:trackrecoPerf}
\end{figure}

 To benchmark the algorithm, we have used a selection of LHCb simulated events of $B_s \to \phi \phi$, being a decay channel of major interest for the LHCb physics programme. The simulations used in this study have been obtained generating $pp$ collisions at centre-of-mass energy $\sqrt{s} = 13.6~{\rm TeV}$ from proton bunches crossing in the 2022-2025 operating conditions of LHC, using \textsc{Pythia}~\cite{PYTHIA} with a specific LHCb configuration~\cite{Gauss2}. Decays of hadronic particles are simulated by \textsc{EvtGen}~\cite{EvtGen}, in which the final-state radiation is generated using \textsc{Photos}~\cite{Photos}. The interaction of the particles with the detector and its response are implemented using \textsc{Geant4}~\cite{GEANT4:2002zbu} as described in Ref.~\cite{Gauss1}. We selected events that leave up to 1000 hits in right-half of the VELO and these have been processed using the described algorithm where a classical solver has been used to solve the system \eqref{eqn:system-of-linear-equation}. To validate the result, the list of active segments is processed into a set of tracks, each one being the set of hits it contains.
The display of an example event and the related reconstructed tracks are shown in Figure \ref{fig:velo_evdisp}. The reconstructed tracks are then validated using the generated, true information and the tracking performances of the algorithm are evaluated in terms of the following metrics:
\begin{itemize}
\item \textbf{Track-finding efficiency} {\boldmath $\epsilon_{\rm track}$}: the fraction of correctly found tracks:

$$
\epsilon_{\rm track} = \frac{N_{\rm track}^{\rm corr}}{N_{\rm gen}^{\rm acc}}\,,
$$
where $N_{\rm track}^{\rm corr}$ is the number of correctly reconstructed tracks and $N_{\rm gen}^{\rm acc}$ is the number of generated particles that are in the acceptance of the detector volume. The acceptance is taken to be the fraction of generated tracks that leave hits in at least 3 detection layers of the right-side of the VELO. For  a track to be considered correctly reconstructed, at least 70\% of the hits of the reconstructed track should be correctly assigned.
\item \textbf{Track-finding fake rate} {\boldmath $f_{\rm fake}$}: the fraction of found tracks that are not associated to a generated particle:

$$
f_{\rm fake} = \frac{N_{\rm track}^{\rm fake}}{N_{\rm track}^{\rm all}}\,,
$$
where $N_{\rm fake} = N_{\rm track}^{\rm all} - N_{\rm track}^{\rm corr}$ is the number of reconstructed tracks that are not associated with a real track. 
\item \textbf{Hit-purity} {\boldmath $e_{\rm pure}$}: the fraction of hits of a track that are correctly assigned to the same generated particle associated with the track:

$$
e_{\rm pure}=\frac{N_{\rm hit}^{\rm correct}}{N_{\rm hit}^{\rm all}}\,,
$$
where $N_{\rm hit}^{\rm all}$ are the total number of hits assigned on a reconstructed track and $N_{\rm hit}^{\rm correct}$ are the correctly assigned hits.

\item  \textbf{Hit-efficiency} {\boldmath$e_{\rm eff}$}: fraction of the hits left by a generated particle that are correctly associated:

$$
e_{\rm eff} = \frac{N_{\rm hit}^{\rm correct}}{N^{\rm gen}_{\rm hit}}\,,
$$
where $N^{\rm gen}_{\rm hit}$ is the total number of hits left by the generated particle.
\end{itemize}
Figure~\ref{fig:trackrecoPerf} shows the performance of the algorithm, in terms of hit-efficiency, hit-purity and track-finding efficiency, as a function of the pseudo-rapidity $\eta$, the momentum $p$ and the transverse momentum $p_\mathrm{T}$ of the generated particles. The algorithm shows an average track-finding efficiency of around $97\%$, with degraded performance at the boundary of the detector acceptance ($\eta$ below $2.5$ and above $4.5$), and for particles with low momentum (below $4\,\mathrm{GeV}$) as their trajectories are affected more by multiple-scattering effects. The hit-efficiency is high (around $98\%$) and shows a constant behaviour, except for the high pseudo-rapidity region where a few hits of high hit-multiplicity tracks are missed with minor impact on the track-finding efficiency. The hit-purity is also high (above $98\%$) with values above $99\%$ in the low pseudo-rapidity region, where tracks have a low hit-multiplicity. The integrated fake tracks rate has been found to be 4.3\%. These results are comparable to the performance of the current state-of-the-art algorithm, Search-by-Triplet~\cite{searchbytriplet}. Nevertheless, the latter has still an edge, indicating that our algorithm can still be further improved with a more refined hyper-parameter search, which goes beyond the purpose of this work.

The algorithm using a classical solver has been also tested on toy events, generated with the model described in Section \ref{sect:toymodel}, containing up to 1000 hits over 26 detection modules and resulted in a perfect (i.e. $100\%$) segment finding efficiency, with a segment purity of 100\% (93\%) when setting the angular parameter to  $\varepsilon = 10^{-9}$ ($10^{-5}$), indicating excellent performance of our new algorithm under ideal conditions.

\subsection{Qiskit implementation}
\begin{figure}[ht]
    \centering
    \includegraphics[width=.7\textwidth]{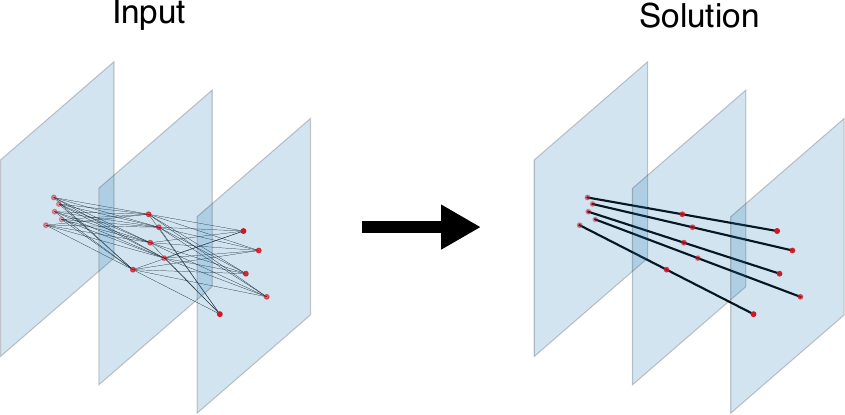}
    \caption{On the left: toy model event with 15 hits left by 5 particles over 3 layers; all the candidate doublets are also shown. On the right: the solution obtained by the Qiskit implementation of the algorithm.}
    \label{fig:qiskit-evdisp}
\end{figure}
To validate the usage of the HHL subroutine, the algorithm has been tested using events generated with a detector toy model, described in Section~\ref{sect:toymodel}. In order to test the implementation using the HHL algorithm, a number of particles ranging from 2 to 5 are generated. For each generated event, the matrix $A$ and the vector $b$ of \eqref{eqn:system-of-linear-equation} are calculated for the list of hits left in the detector modules. The HHL implementation provided by the Qiskit Aqua Python package~\cite{Qiskit} has been used to generate a circuit that solves the system of linear equations. The circuit is then executed on the Qiskit Aer noise-less simulator and results are retrieved by performing a state-vector analysis of the final quantum state.  As an example, the largest (in terms of the number of doublets) reconstructed event is shown in Figure \ref{fig:qiskit-evdisp}. Additionally, circuits are transpiled (optimisation level 3) to fit the \texttt{ibm\_hanoi} 27-qubit r5.11 Falcon quantum processor. This allows us to get realistic circuit sizes for the current implementation of the algorithm on present devices. The number of qubits required, the total depth and number of 2-qubit gates are shown in Table~\ref{tbl:qiskit-results}, together with the size of the input events. Under these conditions, the algorithm was able correctly identify all the tracks in each toy event, with hit-purity $e_\mathrm{pure} = 100\%$  and hit-efficiency $e_\mathrm{eff} = 100\%$.
\begin{table}[ht]
\centering
\begin{tabular}{|r|r|r|r|r|r|}
\hline
\textbf{Layers} & \textbf{Particles} & \textbf{Doublets} & \textbf{Qubits} & \textbf{Depth} & \textbf{2-qubit gates} \\ \hline
3               & 2         & 8                 & 8               & 12 071         & 5 538                  \\ \hline
3               & 3         & 18                & 12              & 1 665 771      & 834 417                \\ \hline
3               & 4         & 32                & 12              & 901 255        & 442 694                \\ \hline
3               & 5         & 50                & 14              & 14 515 229     & 7 107 317              \\ \hline
4               & 2         & 12                & 10              & 185 817        & 93 213                 \\ \hline
4               & 3         & 27                & 12              & 1 714 534      & 840 780                \\ \hline
4               & 4         & 48                & 14              & 14 197 046     & 7 110 044              \\ \hline
\end{tabular}
\caption{The Qiskit implementation has been tested for several event sizes, in terms of the number of layers and particles. For each configuration, the number of qubits and depth of the resulting circuit is reported, as well as the number of 2-qubit gates required on \texttt{ibm\_hanoi}.}
\label{tbl:qiskit-results}
\end{table}
On the downside, the obtained circuit depths range from a few thousands to several millions suggesting that this implementation is not yet in a state to be executed on current circuit-based quantum hardware: the Qiskit implementation of HHL computes of the operator $e^{iAt}$ classically and performs a Pauli decomposition of the result. Only then the resulting unitary matrix is implemented as a quantum circuit. This approach does not take into account the sparsity pattern of the matrix A for the exponentiation and, as a result, the computation has a high classical computational cost and results in very deep circuits, nullifying the advantage provided by HHL. Nevertheless, several theoretical results suggest that, given the high sparsity of the matrix $A$, its condition number behaviour and structure, such an efficient implementation is possible, and it will be investigated in future works~\cite{SparseHamiltonians1,SparseHamiltonians2,SparseHamiltonians3,dsparsehamiltonian}.

\section{Conclusions}

\noindent In this paper, we have presented a new algorithm for the reconstruction of charged particle tracks based on the minimisation of an Ising-like Hamiltonian $\mathcal{H}$. The ground state of $\mathcal{H}$ is obtained using a relaxation procedure that allows turning the problem into a sparse system of linear equations. This approach to the problem is suitable to be implemented on a circuit-based quantum computer using the HHL algorithm to solve the system of linear equations. If all the hypotheses for its efficient implementation are met, this novel approach has the potential to lead to an exponential advantage over classical algorithms.

The studies of the sparsity and condition number scaling show that the algorithm presented in this paper produces sparse systems of linear equations that are very well-conditioned when scaling up the size of the problem, therefore it matches the requirements to achieve an efficient implementation of the HHL algorithm.

Our new Hamiltonian approach has been validated using events coming from simulated $B_s \to \phi\phi$ events in the LHCb vertex detector: the algorithm has shown a performance close to the state of the art, with an average reconstruction efficiency above 96\% and an average fake rate below 5\%, using a classical sparse matrix inversion method. In addition, we have presented a quantum implementation of the same algorithm using the Qiskit library and we have tested it on smaller events generated with a toy model. In all the cases, the correct solution has been retrieved by the algorithm.

An analysis of the size of the resulting quantum circuits clearly shows that the deployment on current quantum hardware is unfeasible at present. This result is strongly limited by the current implementation of Hamiltonian simulation in the QPE subroutine of HHL, which produced extremely deep circuits: this behaviour is expected as the Qiskit implementation of HHL does not make any assumption on the structure of the matrix $A$ and therefore the QPE subroutine becomes extremely expensive. In general, it is hard to achieve efficient simulation if no assumptions are made on the structure of the Hamiltonian. Nevertheless, due to the structure of our matrix $A$, there is strong theoretical support that this should be possible in our use-case~\cite{SparseHamiltonians1,SparseHamiltonians2,SparseHamiltonians3,dsparsehamiltonian}. Future studies on improvements to this section of the algorithm will bring drastic enhancements to the size of the resulting quantum circuits, possibly enabling the execution of the full pipeline on a quantum device.

The HHL algorithm encodes the solution vector as amplitudes of the final state. Therefore, the read-out of the activation state of each segment requires a state tomography, which can be accomplished by using $\mathcal{O}(2^n)$ - $n$ being the number of qubits -  repeated execution of the circuit, voiding the exponential advantage. As a consequence, quantum post-processing of the final state is needed to fully exploit the capabilities of HHL. Quantum clustering techniques could be employed to collect all the active segments and group them into tracks which can then be extracted without full-state tomography. A different approach could exploit the amplitude amplification techniques to increase the amplitude associated with active segments up to a point when the solution can be retrieved with a limited number of shots.

This work opens new possibilities in the application of Quantum Computing to the problem of track reconstruction in HEP. Our new approach, based on the HHL algorithm, has the potential to provide an exponential speed-up over classical counterparts if the conditions discussed above are met. The design of efficient read-out and Hamiltonian simulation subroutines is paramount and will be addressed in future works.

\section{Code availability}
The code associated with this work is available at Ref.~\cite{codeZenodo}. Any update will be made available as a new version at the same Zenodo DOI.

\acknowledgments

We acknowledge the grant support of the Dutch NWO-i funding agency, IBM research Z\"{u}rich, and the Dutch SURF association. We thank in particular Duncan Kampert, Ivano Tavernelli and Ariana Torres for the interesting discussions. We would like to thank the LHCb computing and simulation teams for their support and for producing the simulated LHCb samples used in the paper. The authors would like to thank the LHCb Data Processing \& Analysis (DPA) project colleagues
for supporting this publication and reviewing the work and, in particular, Patrick Koppenburg, Donatella Lucchesi and Eduardo Rodrigues for the valuable discussions.

\appendix
\section{Sparsity studies}
\label{sect:sparsity}

A necessary condition for an efficient implementation of the HHL algorithm is to be able to perform efficient QPE of the unitary operator $e^{iA}$. If the matrix $A$ is sparse, such an efficient implementation should be 
possible\cite{SparseHamiltonians1, SparseHamiltonians2, SparseHamiltonians3,dsparsehamiltonian}. The algorithm presented in this work has been developed under the strict requirement of keeping an high sparsity level when scaling the problem up in size. For a square matrix of size $n$, the sparsity $s$ can be defined as the fraction of zero elements $N_\mathrm{0}$ over the total $n^2$:

$$
s = \frac{N_\mathrm{0}}{n^2}\,.
$$
Within the context of this work, we are interested to study how the sparsity of the matrix $A$ scales with the event size and the impact of the two regularisation term. This study has been done using toy model generated events.\\

\begin{figure}
\begin{subfigure}{0.5\textwidth}
    \centering
    \includegraphics[width=\textwidth]{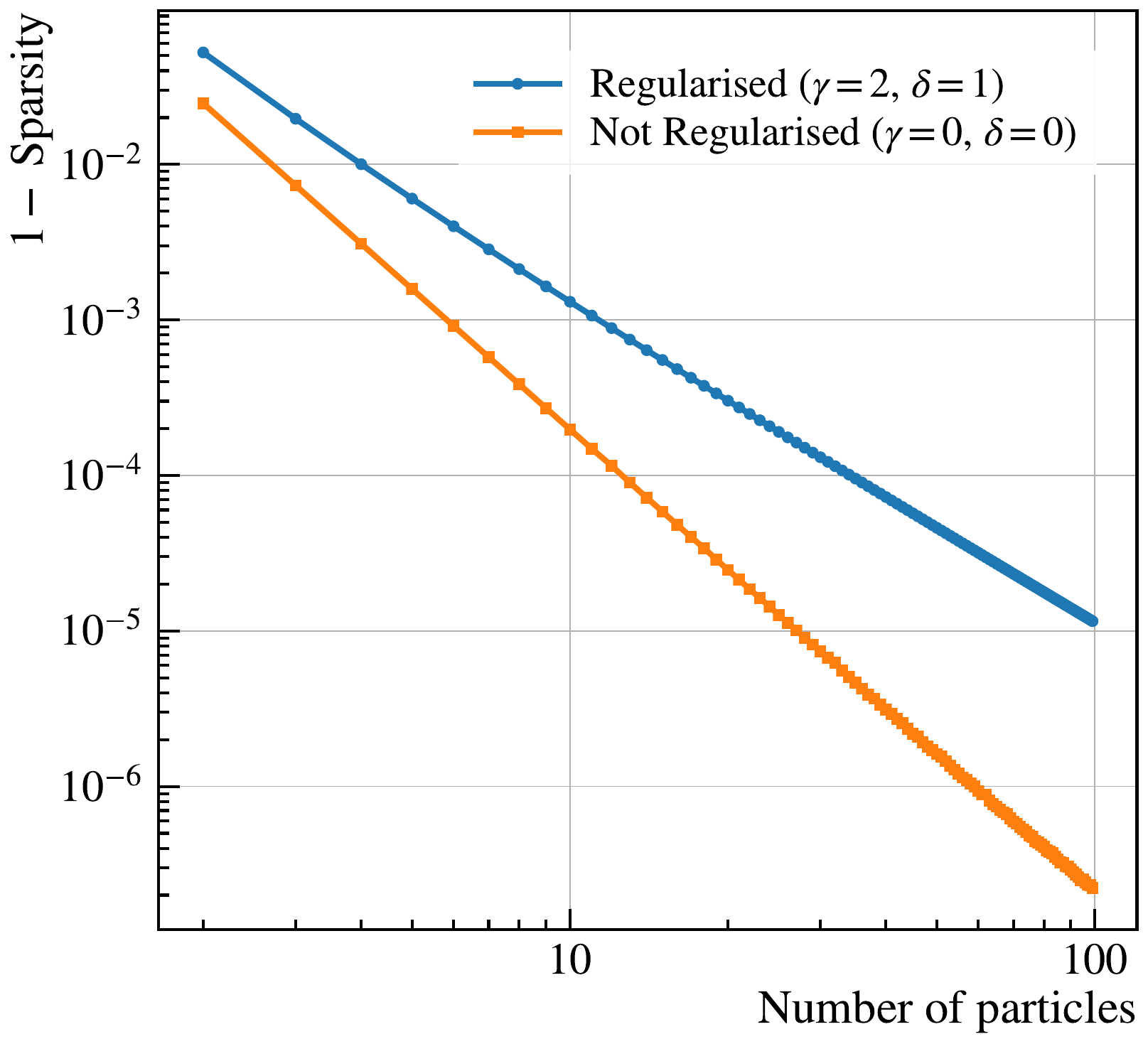}
    \caption{}
    \label{fig:Sparsity_vs_NPart}
\end{subfigure}
\hfill
\begin{subfigure}{0.5\textwidth}
    \centering
    \includegraphics[width=\textwidth]{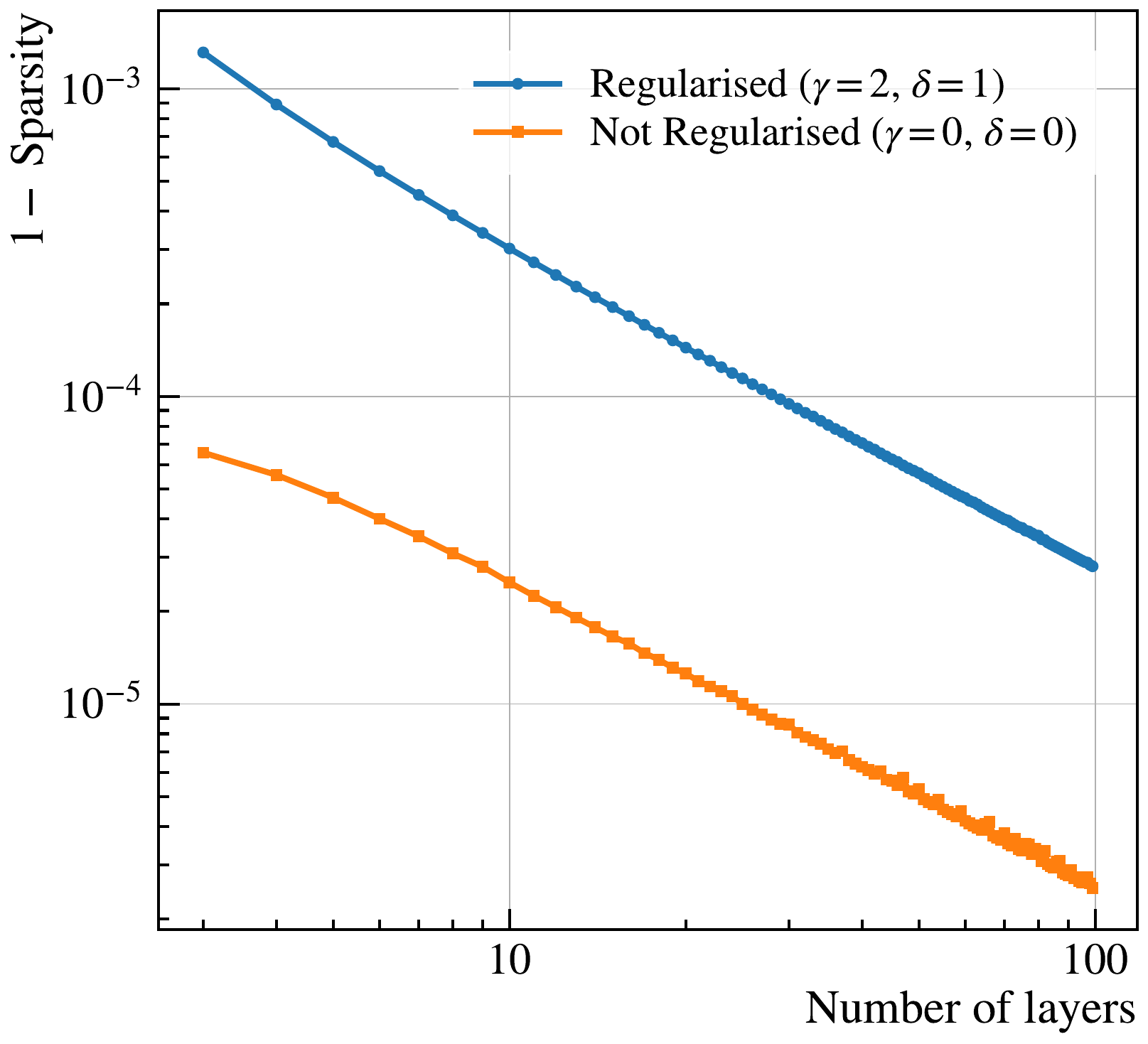}
    \caption{}
    \label{fig:Sparsity_vs_NLayers}
\end{subfigure}
\caption{Sparsity of the $A$ matrix (a) as function of the number of particles in the event (generated with a 10-layer detector) and (b) as function of the number of layers in the detector (fixing the number of particles in the event to 25), for a regularised and not regularised Hamiltonian.}
\end{figure}

\noindent Figures~\ref{fig:Sparsity_vs_NPart} and \ref{fig:Sparsity_vs_NLayers} show that the sparsity of $A$ increases asymptotically to 1 as a power-law, with both the number of particles and the number of detector layers. The regularisation terms do not have a significant impact on this behaviour, as they still preserve the power-law behaviour. This is expected since both the regularisation terms affect the diagonal of the matrix $A$ and therefore they introduce a number of non-zero elements that grows linearly with the matrix size $n$.\\ 

\begin{figure}
    \centering
    \includegraphics[width=.7\textwidth]{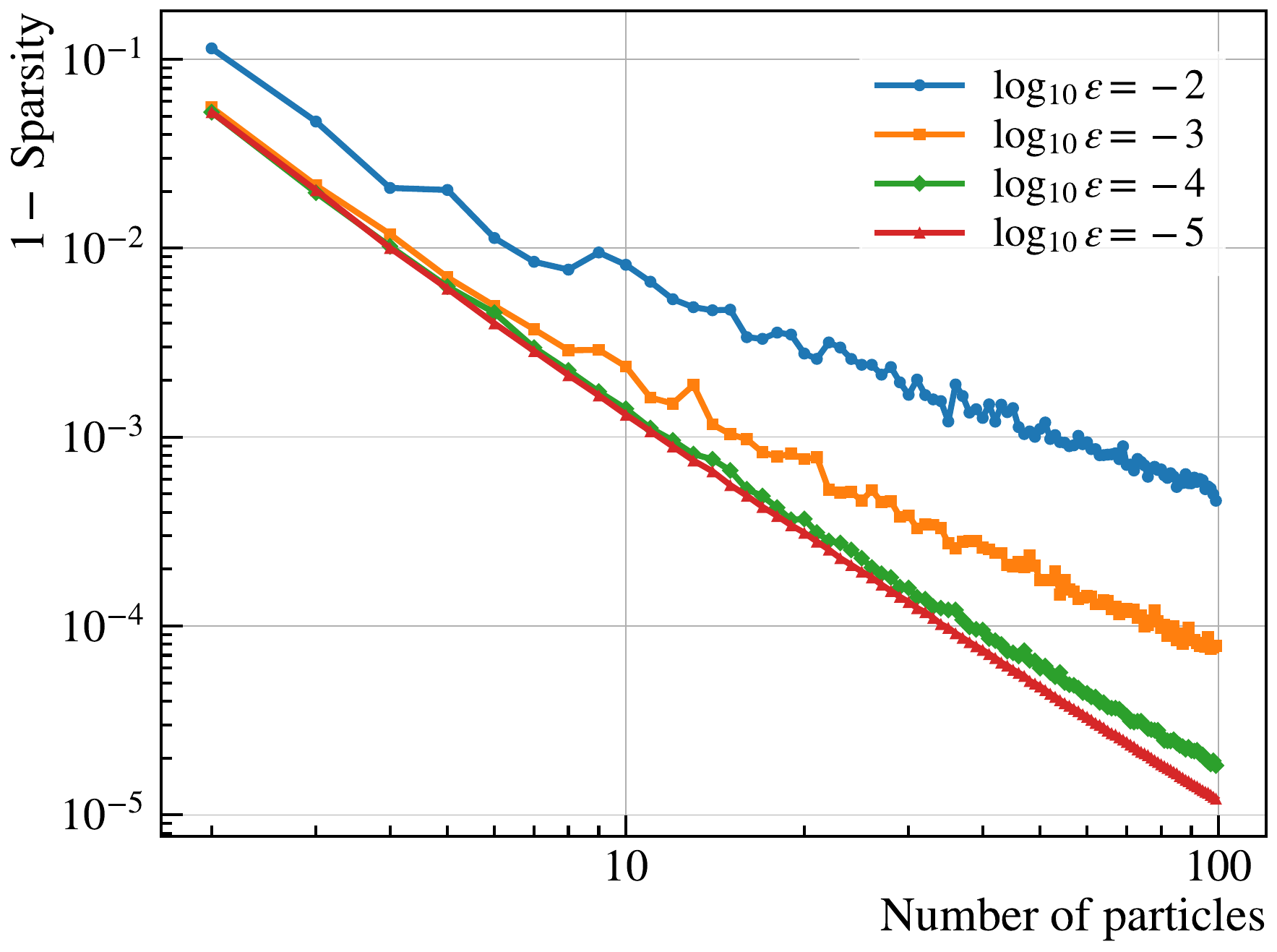}
    \caption{Sparsity of the $A$ matrix as function of the number of particles in the event, for several values of the angular tolerance term $\epsilon$. This has been obtained by processing toy model events with a 10-layer detector. Particles have been generated in a polar angle range of $[0, 1.8^\circ]$ to increase the particle density in the event.}
    \label{fig:Sparsity_vs_Epsilon}
\end{figure}

\noindent On the other end, the angular term contributes to off-diagonal entries of $A$, depending on the angular tolerance $\epsilon$ and the geometrical structure of the event. Therefore, we expect the sparsity to decrease when increasing the angular tolerance and the particle density of the events. This effect has been studied generating toy model events in a smaller polar angle range of $[0, 1.8^\circ]$. Figure~\ref{fig:Sparsity_vs_Epsilon} shows the sparsity versus the number of particles in the event for several values of the angular tolerance $\epsilon$: under this conditions, we can observe a substantial dependence of the sparsity on $\epsilon$, that gets increasingly relevant as the events scale in size. Despite of this, the asymptotic power-law behaviour seems preserved.\\

\section{Condition number studies}
\label{sect:kappa}

The condition number $\kappa$ of the real symmetric matrix $A$ is defined as the ratio between the largest and the smallest (in modulo) eigenvalues of $A$
$$
\kappa(A) = \frac{|\lambda_\mathrm{max}|}{|\lambda_\mathrm{min}|}\, ,
$$
and it plays a major role in the complexity and convergence of several numerical linear algebra methods~\cite{numlinalg}. The HHL algorithm run-time complexity has a quadratic dependence on the condition number that can potentially nullify the exponential advantage when ill-conditioned matrices are involved. In this section, we study the behaviour of the condition number $\kappa(A)$ as a function of the number of particles in the event and layers in the detector.\\

\begin{figure}
    \begin{subfigure}{0.5\textwidth}
        \centering
    \includegraphics[width=\textwidth]{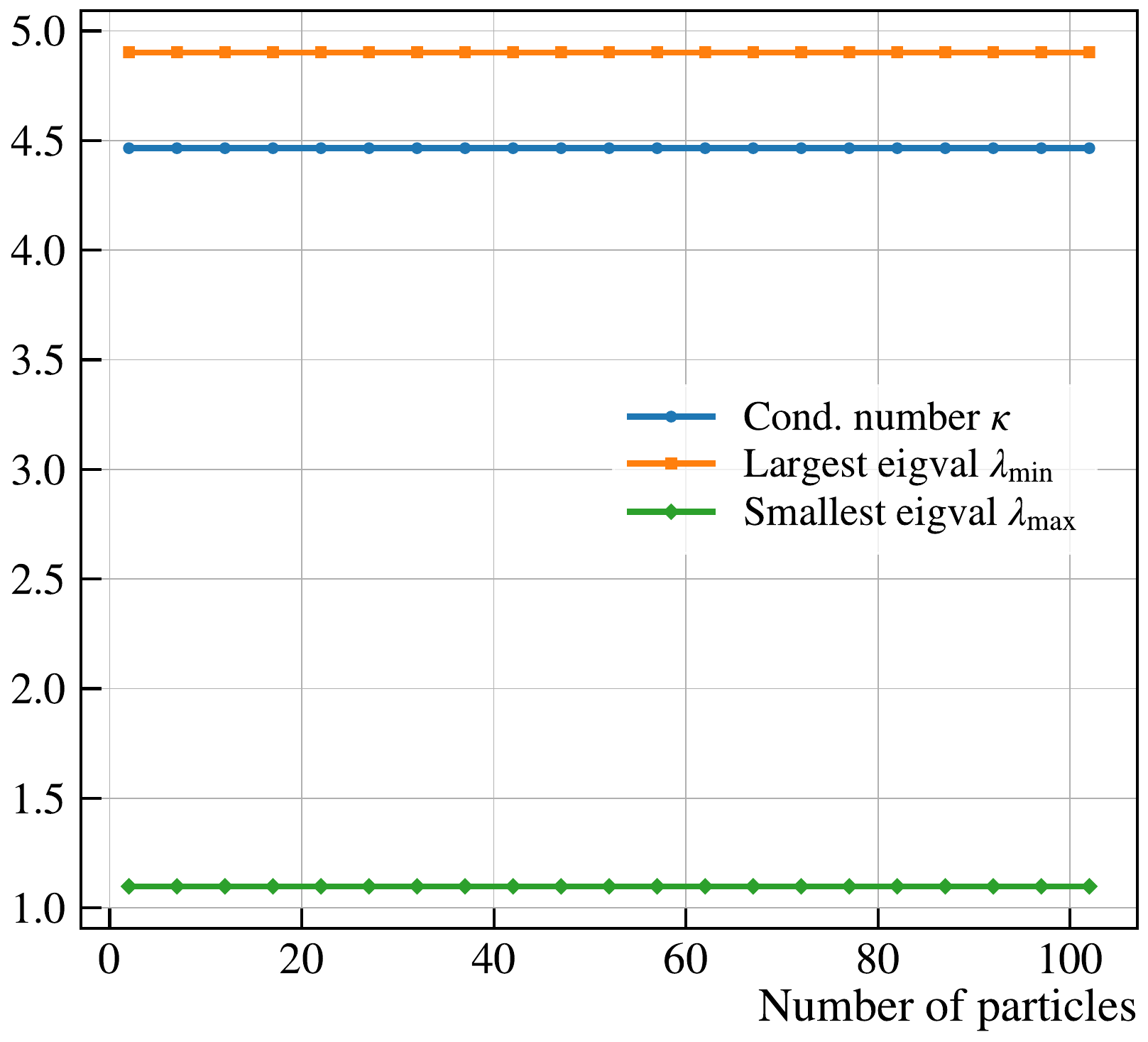}
    \caption{}
    \label{fig:kappa_vs_part}
    \end{subfigure}
    \hfill
    \begin{subfigure}{0.5\textwidth}
    \centering
    \includegraphics[width=\textwidth]{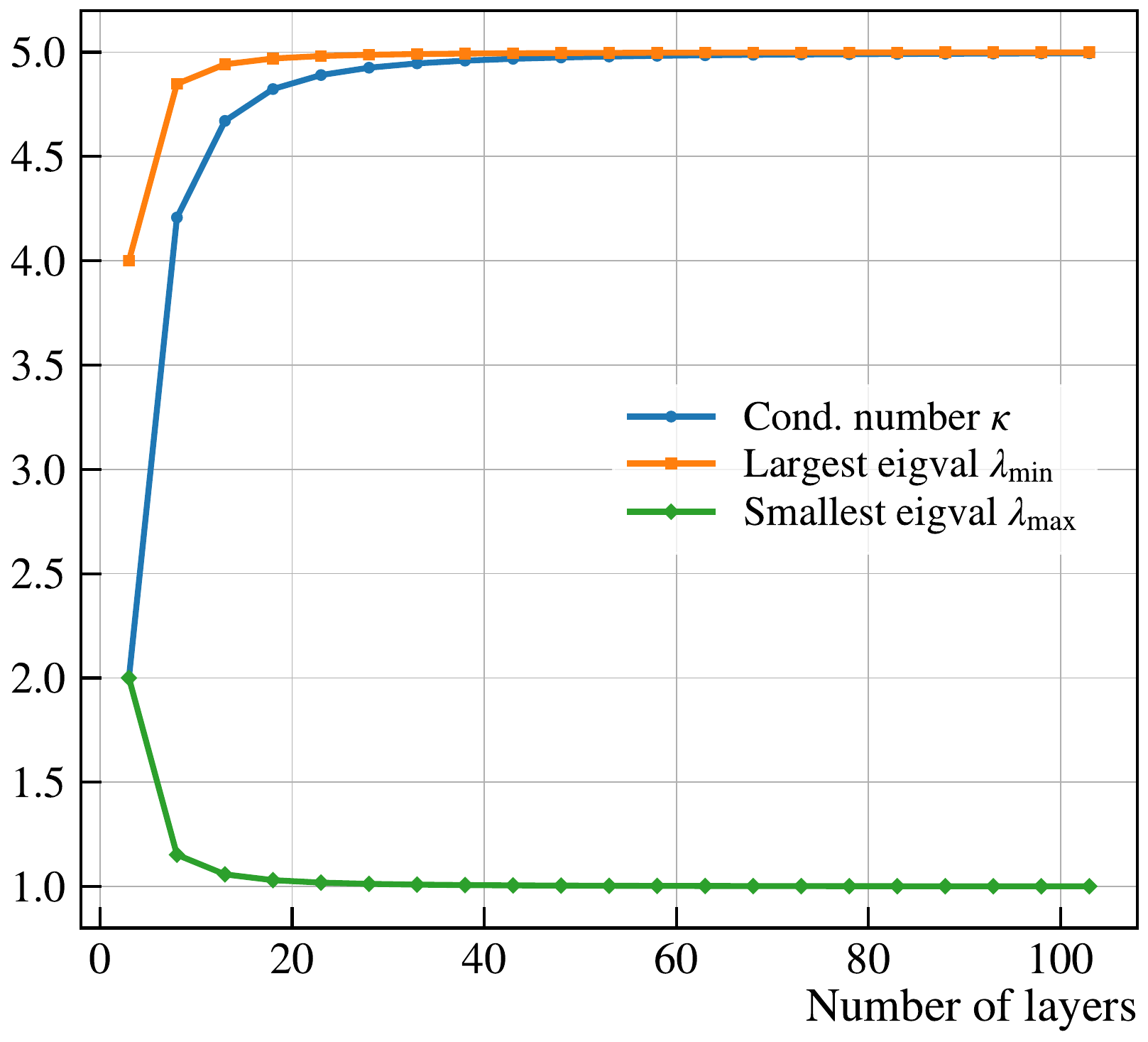}
    \caption{}
    \label{fig:kappa_vs_layers}
    \end{subfigure}

    \caption{Condition number $\kappa$, largest eigenvalue $\lambda_\mathrm{max}$ and smallest eigenvalue $\lambda_\mathrm{min}$ of $A$ (a) as a function of the number of particles  in toy model event (generated with a 10-layer detector) and  (b) as a function of the number of layers  in the toy model detector (the number of particles in the event is fixed to 25).}
    
\end{figure}

\noindent Figure~\ref{fig:kappa_vs_part} clearly shows that the largest and smallest eigenvalues have no dependence on the number of particles in the event, thus the condition number is constant. On the other hand, studying the behaviour as a function of the number of layers, Figure~\ref{fig:kappa_vs_layers} shows that the largest(smallest) eigenvalue has asymptotically increases(decreases) towards 5(1). Therefore, the condition number increases with the number of layers but it can be safely bounded from above: $\kappa < 5$. The bound holds irrespective of the event/detector size, allowing us to ignore the dependence of the run-time complexity on $\kappa$. This property makes our approach to track reconstruction an excellent candidate for an HHL application.

\FloatBarrier
\bibliographystyle{jhep}
\bibliography{bibliography}
\end{document}